\documentclass[superscriptaddress,12pt]{revtex4}
\usepackage{dcolumn}
\usepackage{bm}
\usepackage{graphicx}
\usepackage{amsfonts}
\usepackage{amsmath}
\usepackage{amssymb}
\usepackage[T1]{fontenc}
\usepackage[frenchb]{babel}

\begin{document}

\title{Signature of Magnetic Monopole and Dirac string Dynamics in Spin Ice}

\author{L. D. C. Jaubert}
\affiliation{Universit\'e de Lyon, Laboratoire de Physique, \'Ecole Normale Sup\'erieure de Lyon, 46 All\'ee d'Italie, 69364 Lyon Cedex 07, France.}

\author{P. C. W. Holdsworth}
\affiliation{Universit\'e de Lyon, Laboratoire de Physique, \'Ecole Normale Sup\'erieure de Lyon, 46 All\'ee d'Italie, 69364 Lyon Cedex 07, France.}

\maketitle

\textbf{Magnetic monopoles have eluded experimental detection since their prediction nearly a century ago by Dirac~\cite{Di31}. Recently it has been shown that classical analogues of these enigmatic particles occur as excitations out of the topological ground state of a model magnetic system, dipolar spin ice~\cite{Ca08}. These quasi-particle excitations do not require a modification of Maxwell's equations, but they do interact via Coulombs law and are of magnetic origin. In this paper we present an experimentally measurable signature of monopole dynamics and show that magnetic relaxation measurements in the spin ice material $Dy_{2}Ti_{2}O_{7}$~[3] can be interpreted entirely in terms of the diffusive motion of monopoles in the grand canonical ensemble, constrained by a network of ``Dirac strings'' filling the quasi-particle vacuum. In a magnetic field the topology of the network prevents charge flow in the steady state, but there is a monopole density gradient near the surface of an open system.}\\\\

\begin{figure}[ht]
\includegraphics[width=10cm]{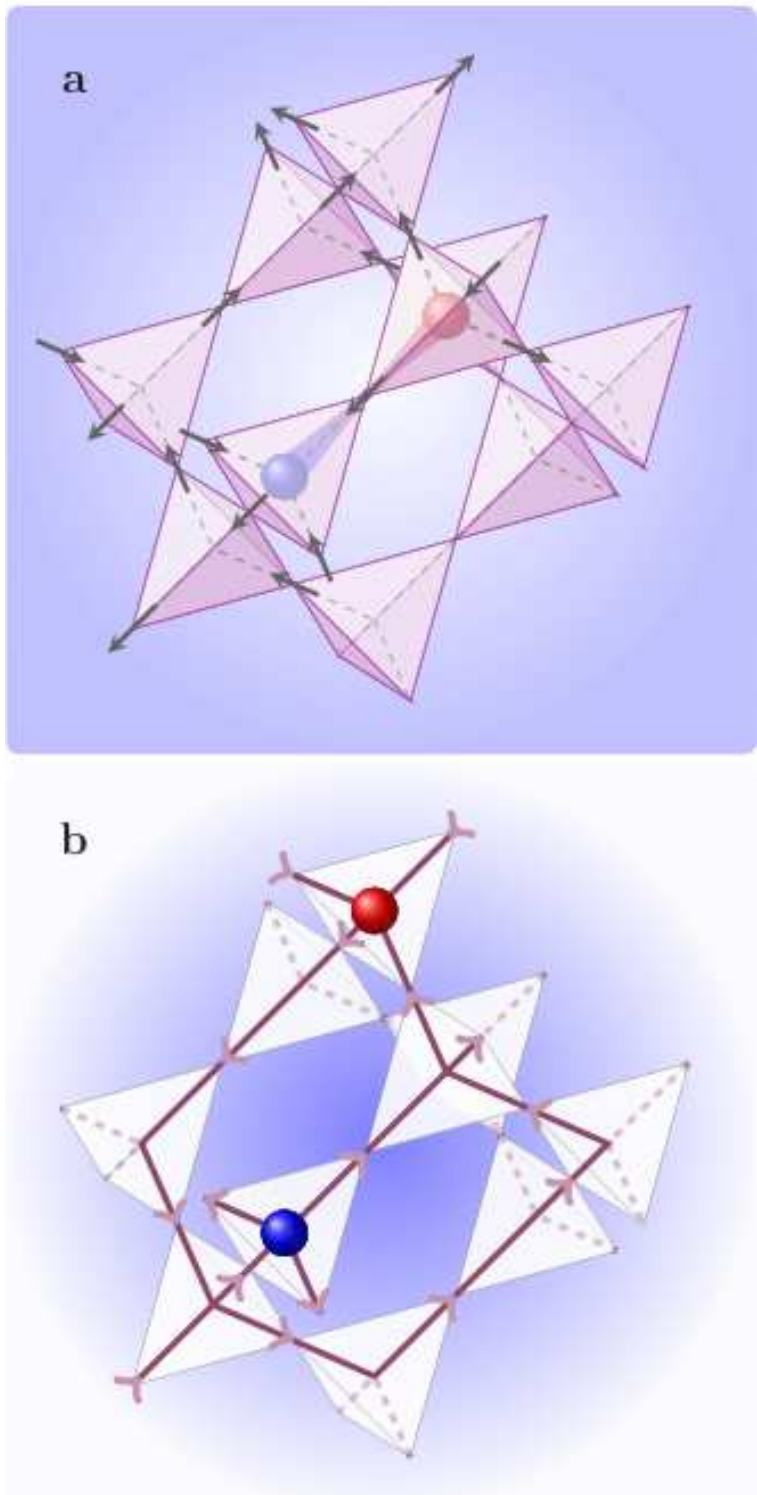}
\caption{\textbf{Spin Ice structure and emergence of monopoles.} a) The magnetic ions ($Ho^{3+}$ or $Dy^{3+}$) lie on the sites of the pyrochlore lattice and are constrained to the bonds of the duel diamond lattice (dashed lines). Local topological excitations 3 in - 1 out or 3 out - 1 in correspond to magnetic monopoles with positive (blue sphere) or negative (red sphere) charges respectively. b) The diamond lattice provides the skeleton for the network of Dirac strings with the position of the monopole restricted to the vertices. The orientation of the Dirac strings shows the direction of the local field lines in $\mathbf{H}$.}
\label{fig1}
\end{figure}

Spin ice systems~\cite{Ha97,Br01,Br04} such as $Dy_{2}Ti_{2}O_{7}$ and $Ho_{2}Ti_{2}O_{7}$, can be described by a corner sharing network of tetrahedra forming a pyrochlore lattice of localized magnetic moments, as shown in figure~\ref{fig1}a. The pairwise interaction is made up of both exchange and dipolar terms
\begin{eqnarray}
\mathcal{H}\;=\;Jm^{2}\,\sum_{\left<i,j\right>}\mathbf{S}_{i}\cdot\mathbf{S}_{j}\;+\;
Dm^{2}\,\sum_{\left<i,j\right>}\left[
\frac{\mathbf{S}_{i}\cdot\mathbf{S}_{j}}{\left|\mathbf{r}_{ij}\right|^{3}}-
\frac{3\left(\mathbf{S}_{i}\cdot\mathbf{r}_{ij}\right)\left(\mathbf{S}_{j}\cdot\mathbf{r}_{ij}\right)}{\left|\mathbf{r}_{ij}\right|^{5}}\right]
\label{eq1}
\end{eqnarray}
where the rare earth ions carry a moment of ten Bohr magnetons, $m\approx 10\mu_{B}$  and where $\mathbf{S}_{i}$ is a spin of unit length. The coupling constants are on the $1\,K$ energy scale; for example for $Dy_{2}Ti_{2}0_{7}$ $|J|m^{2}\approx3.72\,K$ and $Dm^{2}\approx1.41\,K$ ~\cite{dH00}. These energy scales are 100 times smaller than the crystal field terms~\cite{Eh03} that confine the spins along the axis joining the centres of two adjoining tetrahedra. As a result, on the $1\,K$ energy scale the moments behave as Ising spins along this axis. Remarkably, within this Ising description the long ranged dipolar interactions are almost perfectly screened~\cite{dH00,Is05} at low temperature, with the result that the low energy properties are almost identical to those of an effective frustrated nearest neighbour model with antiferromagnetic interactions~\cite{An56} of strength $J_{eff}=(5D-J)m^{2}/3$ . This is equivalent to Pauling's model for proton disorder in the cubic phase of ice~\cite{Pa35}, which has extensive ground state entropy and violates the third law of thermodynamics~\cite{Br04}. It successfully reproduces the thermodynamic behaviour of both ice~\cite{Gi36} and spin ice~\cite{Ra99} and describes the microscopic properties of the latter to a good approximation. The extensive set of spin ice states satisfy the Bernal - Fowler ice rules~\cite{Be33}; a $3d$ analogue of the 6 vertex model with topological constraint consisting of two spins pointing into and two out of each tetrahedron (2 in - 2 out), as shown in figure~\ref{fig1}a.  Flipping one spin breaks the constraint leaving neighbouring tetrahedra with 3 in - 1 out and with 3 out - 1 in, which constitute a pair of topological defects.  Within the nearest neighbour model, creation of the defect pair costs energy $4\,J_{eff}$, while further spin flips can move the defects at zero energy cost. It has recently been shown~\cite{Ca08} that including the full dipolar Hamiltonian of equation (1) leads to an effective Coulombic interaction between the topological defects separated by distance $r$ , $\mu_{0}q_{i}q_{j}/4\pi r$, where $\mu_{0}$  is the permeability of free space, $q_{i}=\pm q=\pm 2m/a$, and $a$ is the distance between two vertices of the diamond lattice (see figure~\ref{fig1}); that is, to a Coulomb gas of magnetic monopoles. Standard electromagnetic theory does allow for such excitations~\cite{Ja99}, which correspond to divergences in the magnetic intensity $\mathbf{H}$, or magnetic moment $\mathbf{M}$, rather than in the magnetic induction: $\mathbf{\nabla}\cdot\mathbf{B}=\mathbf{\nabla}\cdot(\mathbf{H}+\mathbf{M})=0$. On all length scales above the atomic scale, a 3 in - 1 out defect appears to be a local sink in the magnetic moment and therefore as a source of field lines in $\mathbf{H}$ . It can lower its energy by moving in the direction of an external field and therefore carries a positive magnetic charge~\cite{Ry05}. What is remarkable about spin ice is that it allows for the deconfinement of these effective magnetic charges so that they occur in the bulk of the material on all scales, rather than just at the surfaces within a coarse grained description~\cite{Ja99}. A two dimensional equivalent may exist in artificial spin ice, constituting arrays of nanoscale magnets~\cite{Wa06}.\\

\begin{figure}[ht]
\includegraphics[width=15cm]{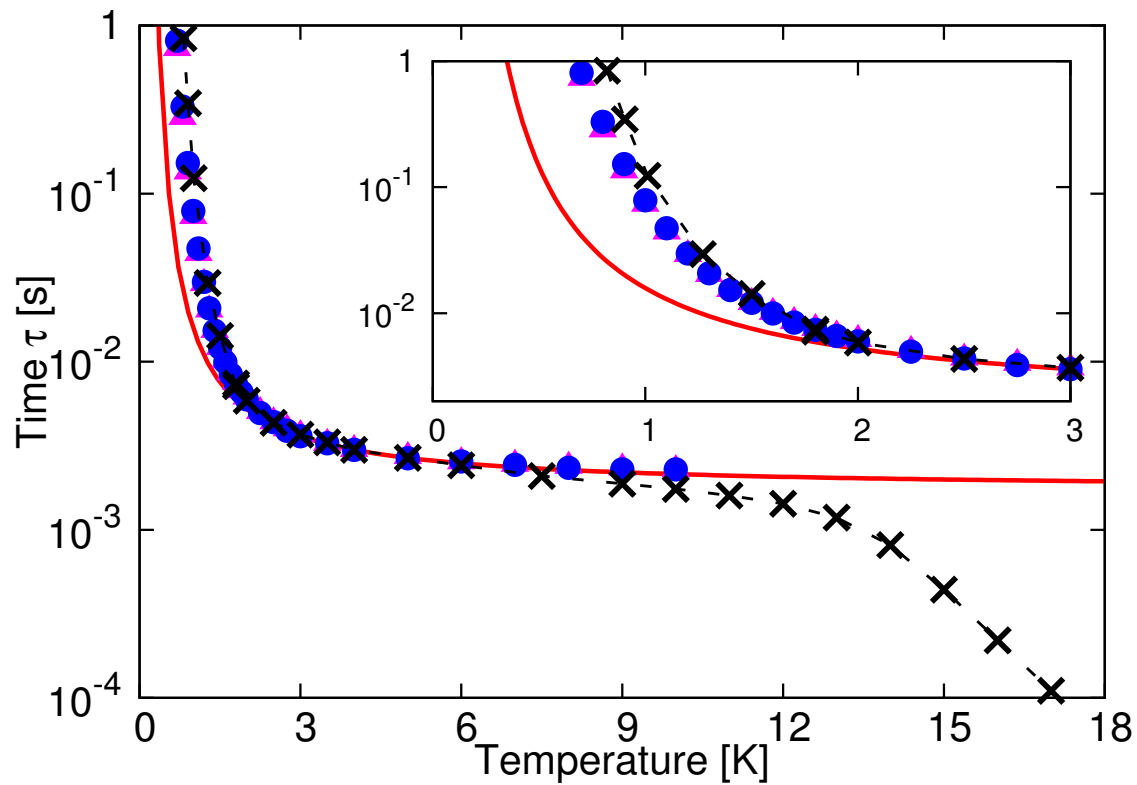}
\caption{\textbf{Relaxation time scales $\tau$ in $\mathbf{ Dy_{2}Ti_{2}O_{7}}$: experiment and simulation.} The experimental data ($\mathbf{\times}$) are from Snyder \& al. [3]. The Arrhenius law (red line) represents the free diffusion of topological defects in the nearest neighbour model. The relaxation time scale of the Dirac string network driven by Metropolis dynamics of magnetic monopoles has been obtained for fixed chemical potential (pink {\Large $\blacktriangle$}) and with $\mu$ varying slowly to match the defect concentration in dipolar spin ice (blue {\huge $\bullet$}). The temperature scale is fixed without any free parameters. Inset: Same data shown in the low temperature region.}
\label{fig2}
\end{figure}

Given the accessibility of these magnetic quasi-particles, the development of an experimental signature is of vital importance and interest.  The ``Stanford'' superconducting coil experiment~\cite{Ca08,Ca82} could in principle detect the passage of a single magnetic quasi-particle, but this seems highly unlikely given that the charges have no mass and therefore have diffusive, rather than Newtonian dynamics. A more promising starting point is therefore to look for a monopole signal from magnetic relaxation of a macroscopic sample~\cite{Sn04,Eh03,Ma00}. The general dynamic behaviour of spin ice is illustrated in figure~\ref{fig2} by the magnetic relaxation time, as a function of temperature for $Dy_{2}Ti_{2}O_{7}$~\cite{Sn04}, taken from bulk susceptibility measurements. The energy scales discussed above give rise to different regimes: the time scale increases in the thermally activated high temperature regime, entering a quasi-plateau region below $12\,K$ associated with quantum tunnelling processes~\cite{Eh03}, before experiencing a sharp upturn below $2\,K$. The spins are Ising like below $12\,K$ and the configuration evolves by quantum tunnelling through the crystal field barrier, while above this temperature higher crystal field levels are populated and the time scale drops dramatically. The quantum tunnelling plateau regime can therefore be well represented by an Ising system with stochastic single spin dynamics and hence should be dominated by the creation and propagation of monopole objects. This is illustrated, in a first approximation, by comparing the data with an Arrhenius law $\tau=\tau_{0}\,\exp(2\,J_{eff}/k_{B}T)$, as shown by the red curve in figure~\ref{fig2}. The time scale $\tau_{0}$ is fixed by fitting to the experimental time at $4\,K$ with $J_{eff}=1.11\,K$, the value estimated for $Dy_{2}Ti_{2}0_{7}$~\cite{dH00}. $2\,J_{eff}$ is the energy cost of a single, free topological defect in the nearest neighbour approximation and is half that for a single spin flip. The calculation fits the data over the low temperature part of the quasi-plateau region, where one expects a significant defect concentration without any double defects (4-in or 4-out), and gives surprisingly good qualitative agreement at lower temperature, as the concentration decreases. Although still in the tunnelling regime, the plateau region corresponds to high temperature for the effective Ising system. Good agreement here provides a stringent test and any theory not fitting must be discarded. The above expression clearly does a good job, allowing us to equate $\tau_{0}$ with the microscopic tunnelling time. This test therefore already provides very strong evidence for the fractionalization of magnetic charge~\cite{Ca08} and the diffusion of unconfined particles. However, this (or any other) Arrhenius function ultimately fails, underestimating the time scale at very low temperature: while it is possible to fit the data reasonably below $2\,K$ by a single exponential function by varying the barrier height, simultaneous agreement along the plateau and at lower temperature is impossible. The role of the missing Coulomb interaction is therefore clear: although non-confining it must considerably increase the relaxation time scale by modifying the defect concentration and slowing down diffusion through the creation of locally bound pairs.\\

We have tested this idea by directly simulating a Coulomb gas of magnetically charged particles (monopoles), in the grand canonical ensemble, occupying the sites of the diamond lattice. The magnetic charge is taken as $q_{i}=\pm q$. In the grand canonical ensemble the chemical potential is an independent variable, whose value in the corresponding magnetic experiment is unknown. In a first series of simulations we have estimated it numerically by calculating the difference between the Coulomb energy gained by creating a pair of neighbouring magnetic monopoles and that required to produce a pair of topological defects in the dipolar spin ice model, with parameters taken from reference~\cite{dH00}, giving a configurationally averaged estimate $\mu/k_{B}=8.92\,K$. In a second series of simulations $\mu$ was taken as the value required to reproduce the same defect concentration as in a simulation of dipolar spin ice at temperature $T$. Here $\mu$ varied by 3\% only, with the same mean value as in the first series, showing that our procedure is consistent. The chemical potential used is thus \textit{not} a free parameter. As the Coulomb interaction is long ranged, we treat a finite system using the Ewald summation method~\cite{dL80,Fr02}.  The monopoles hop between nearest neighbour sites via the Metropolis Monte Carlo algorithm, giving diffusive dynamics, but with a further local constraint: in the spin model a 3 in  - 1 out topological defect can move at low energy cost by flipping one of the three in spins, the direction of the out spin being barred by an energy barrier of $8\,J_{eff}$. An isolated monopole can therefore hop to 3 out of 4 of its nearest neighbour sites only, dictated by an oriented network of constrained trajectories similar to the ensemble of classical ``Dirac string''~\cite{Ca08} of overturned dipoles~\cite{Ja99}. The positively charged monopoles move in one sense along the network while the negative charges move in the opposite direction (see figure~\ref{fig1}b). The network is dynamically re-arranged through the evolution of the monopole configuration. The vacuum for monopoles in spin ice thus has an internal structure; the Dirac strings which, in the absence of monopoles, satisfy the ice rules at each vertex. This structure is manifest in the dynamics and influences the resulting time scales. In fact the characteristic time scale that we compare with experiment comes from the evolution of the network of Dirac strings rather than from the monopoles themselves. Indeed, the monopole autocorrelation time, as extracted from the monopole density - density correlation function~\cite{Ba03} turns out to be small for this range of temperature. We locally define the string network by an integer $\sigma=\pm 1$, giving the orientation of the Dirac string along each bond of the diamond lattice and define the autocorrelation function
\begin{eqnarray}
C(t)\;=\;\frac{1}{N}\sum_{i}\sigma_{i}(t)\sigma_{i}(0),
\end{eqnarray}
where $t$ is the Metropolis time and $N$ is the number of bonds (up to $N\approx 25000$). For the initial conditions we take an ordered network with no monopoles, which we let evolve at temperature $T$ until an equilibrium configuration is attained. This defines $t=0$. $C(t)$ decays almost exponentially, with characteristic time, $\tau$, that varies with temperature. To avoid initial transient effects we define $\tau$ such that $C(\tau)=0.8$. The time is re-set to zero when $C(t)$ decays beyond 0.01 and the process is repeated many times to give the configurationally averaged decay time. In figure~\ref{fig2} we compare our simulations with the experimental data of reference~\cite{Sn04}. The Metropolis time is again scaled to the experimental time at $4\,K$ and there is again no scale factor on the temperature axis. Data for fixed chemical potential are shown by the pink triangles, while data with $\mu$ varying are shown by the blue circles. There is a quantitative evolution of the simulation data compared to the nearest neighbour spin ice model. Agreement between the experimental and numerical data now looks excellent, showing clearly that the experimental relaxation is due to the creation and proliferation of quasi-particle excitations that resemble classical monopoles in the magnetic intensity $\mathbf{H}$. As the temperature increases, towards the end of the plateau region, a small systematic difference occurs. This is because the spin system can access double defects at finite energy cost, while this state corresponds to two like charges superimposed on the same site, which is excluded by the Coulomb interaction. The inset of figure~\ref{fig2} shows results at low temperature illustrating in detail the extent of the improvement in comparison with experiment. Allowing the variation of $\mu$ provides a further evolution towards the experimental data, compared with that for fixed $\mu$ and the blue circles represent our best numerical results. We now have quantitative agreement between experiment and theory down to low temperature, showing that the Coulomb interactions are responsible for the non-Arrhenius temperature dependence of the relaxation time scales. Differences remain at this level of comparison below $1\,K$ but to go further would require an even more detailed modelling of spin ice~\cite{Ya08} as well as complementary experimental measurements.\\

\begin{figure}[ht]
\includegraphics[width=15cm]{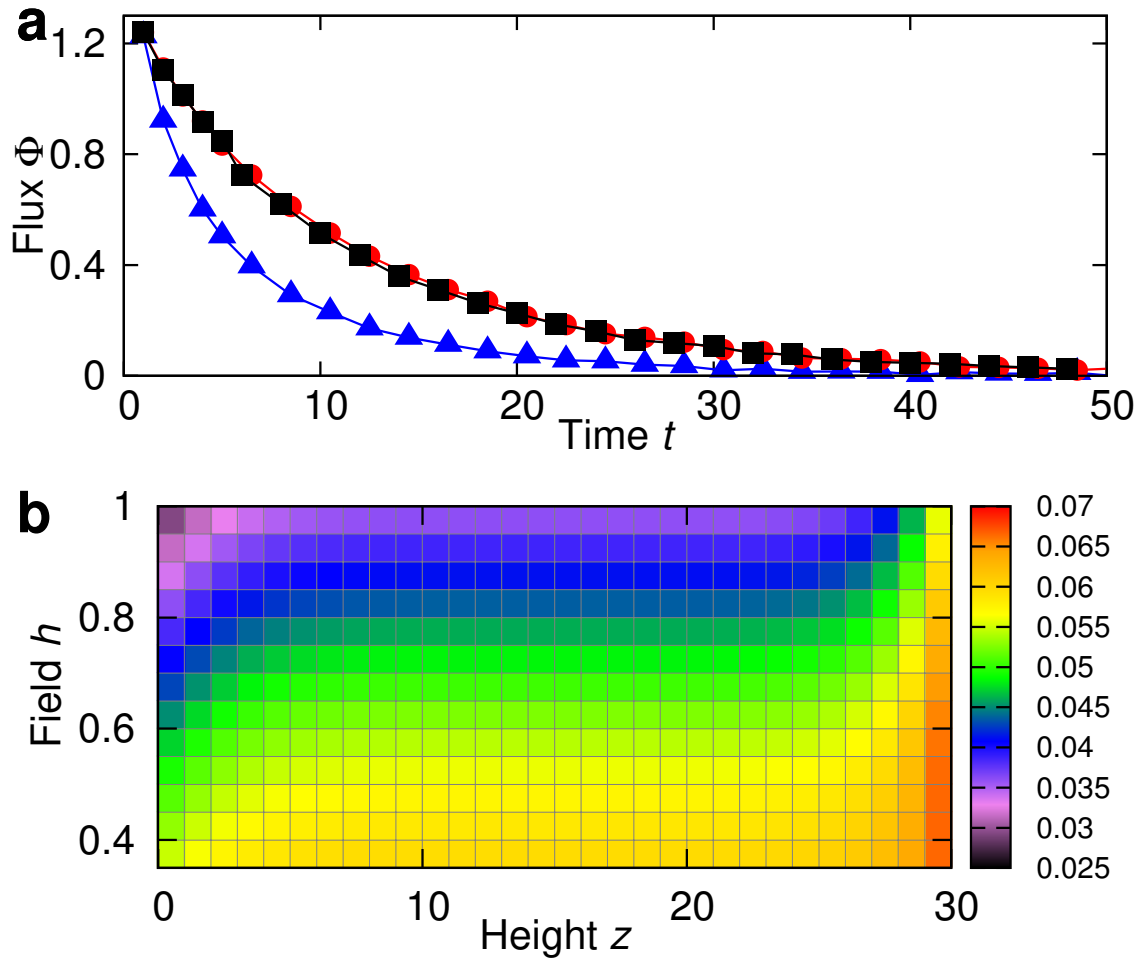}
\caption{\textbf{Monopole density profile.} a) A magnetic field $h$ is applied at $t=0$ along the $z$-axis ([100] direction). We display the transient flux of positive charges, $\Phi$, passing through a plane perpendicular to the field, as a function of Metropolis time, $t$. The simulations are obtained using the nearest neighbour spin ice model with periodic boundary conditions ({\large $\blacksquare$}) and open boundaries, with current measured either at the surface (blue {\Large $\blacktriangle$}) or in the bulk (red {\huge $\bullet$}). b) Density of positive defects in the horizontal planes for $T=1\,K$, as a function of $z$ and $h$ (in units of $k_{B}T/m\mu_{0}$).}
\label{fig3}
\end{figure}

Finally we consider the response of monopoles to an external magnetic field, $h$, placed along one of the [100] directions. Applying such a field to a system for closed circuit geometry (periodic boundaries), one might expect the development of a monopole current in the steady state~\cite{Ry05}. This is not the case, at least for the nearest neighbour model, where we find that a transient current decays rapidly to zero (see figure~\ref{fig3}a). The passage of a positive charge in the direction of the field re-organises the network of strings, leaving a wake behind it that can be followed either by a negative charge, or by a positive one moving against the field, with the result that the current stops. This is a dynamic rather than static effect and is not related to confinement of monopole pairs by the background magnetization~\cite{Ca08}. Reducing the temperature at finite field, the magnetization saturates around a critical temperature; a vestige of the Kasteleyn transition~\cite{Ja08}, which is unique to topologically constrained systems. Confinement occurs here, as the Zeeman energy outweighs the entropy gain of free monopoles. The transient currents suggest the development of charge separation in an open system. This is indeed the case despite the fact that monopole numbers are not conserved at open boundaries. In figure~\ref{fig3}b we show the profile of positive charge density across a sample of size $L$, with open boundaries, for varying fields. There is a clear build up of charge over a band of 4-5 lattice spacings, although including long range interactions may lead to a quantitative change in this value.  As the ratio $T/h$ and the monopole density go to zero the band narrows and the system forms a conventional layer of magnetic surface charge as one expects for any magnetically ordered system~\cite{Ja99}. In the absence of topological defects the magnetization is conserved from one layer to another, so that a charge density profile manifests itself as a magnetization profile. The data here suggest charge build up in a layer several nanometres thick, making it in principle a measurable effect.

\noindent\textbf{Acknowledgements} We thank S.T. Bramwell, C. Castelnovo, J.T. Chalker, M. Clusel, M.J.P. Gingras and R. Melko for useful discussions and P. Schiffer for providing the data from reference [3]. We acknowledge financial support from the European Science Foundation PESC/RNP/HFM.\\\\
\textbf{Author Contributions} The authors contributed equally to the manuscript.\\\\
\textbf{Author Information} Correspondence should be addressed to L.J. (e-mail: ludovic.jaubert@ens-lyon.org).\\

\end{document}